\documentclass[pra,twocolumn,showpacs,amsmath,amssymb,superscriptaddress,eqsecnum]{revtex4-1}
\usepackage{graphicx,bm,color,mathptmx,hyperref}
\newcommand{\op}[1]{\hat{#1}}

\newcommand{\pd}[2]{\frac{\partial #1}{\partial #2}} 
\newcommand{\pdd}[2]{\frac{\partial^2 #1}{\partial #2^2}} 
\usepackage{soul}
\definecolor{lightblue}{RGB}{185,210,248}
\sethlcolor{lightblue}

\begin{document}

\title{Radial quantum number of Laguerre-Gauss modes}

\author{E. Karimi}
\affiliation{ Department of Physics, University of Ottawa, 150 Louis
  Pasteur, Ottawa, ON K1N 6N5 Canada}

\author{R. W. Boyd}
\affiliation{ Department of Physics, University of Ottawa, 150 Louis
  Pasteur, Ottawa, ON K1N 6N5 Canada}
 
\author{P. de la Hoz} 
\affiliation{Departamento de \'Optica, Facultad
  de F\'{\i}sica, Universidad Complutense, 28040~Madrid, Spain}

\author{H. de Guise}
\affiliation{Department of Physics, Lakehead University, 
Thunder Bay, ON P7B 5E1, Canada}

\author{J. \v{R}eh\'{a}\v{c}ek}
\affiliation{Department of Optics,
Palack\'{y} University, 17. listopadu 12,
746 01 Olomouc, Czech Republic}

\author{Z. Hradil}
\affiliation{Department of Optics,
Palack\'{y} University, 17. listopadu 12,
746 01 Olomouc, Czech Republic}

\author{A.~Aiello}
\affiliation{Max-Planck-Institut f\"ur die Physik des Lichts, 
G\"{u}nther-Scharowsky-Stra{\ss}e 1, Bau 24, 
91058 Erlangen,  Germany}
\affiliation{Department f\"{u}r Physik, Universit\"{a}t Erlangen-N\"{u}rnberg,
Staudtstra{\ss}e 7, Bau 2, 91058 Erlangen, Germany}

\author{G.~Leuchs}
\affiliation{Max-Planck-Institut f\"ur die Physik des Lichts, 
G\"{u}nther-Scharowsky-Stra{\ss}e 1, Bau 24, 
91058 Erlangen,  Germany}
\affiliation{Department f\"{u}r Physik, Universit\"{a}t Erlangen-N\"{u}rnberg,
Staudtstra{\ss}e 7, Bau 2, 91058 Erlangen, Germany}
\affiliation{ Department of Physics, University of Ottawa, 150 Louis
  Pasteur, Ottawa, ON K1N 6N5 Canada}

\author{L.~L.~S\'anchez-Soto} 
\affiliation{Max-Planck-Institut f\"ur die Physik des Lichts,
  G\"{u}nther-Scharowsky-Stra{\ss}e 1, Bau 24, 91058 Erlangen,
  Germany} 
\affiliation{Department f\"{u}r Physik, Universit\"{a}t
  Erlangen-N\"{u}rnberg, Staudtstra{\ss}e 7, Bau 2, 91058 Erlangen,
  Germany}
\affiliation{Departamento de \'Optica,
  Facultad de F\'{\i}sica, Universidad Complutense, 28040~Madrid,
  Spain}

\begin{abstract}
  We introduce an operator linked with the radial index in the
  Laguerre-Gauss modes of a two-dimensional harmonic oscillator in
  cylindrical coordinates. We discuss ladder operators for this
  variable, and confirm that they obey the commutation relations of
  the su(1,1) algebra. Using this fact, we examine how basic quantum
  optical concepts can be recast in terms of radial modes.
\end{abstract}


\date{\today}

\maketitle

\section{Introduction}
 
An optical vortex is a light field exhibiting a pure screw phase
dislocation along the propagation axis; i.e., an azimuthal phase
dependence $\exp ( i \ell \varphi)$. The integer $\ell$ plays the role
of a topological charge: the phase changes its value by $\ell$ cycles
of $2 \pi$ in any closed circuit about the axis, while the amplitude
is zero there~\cite{Torres:2011vn}.

One of the most intriguing properties of vortices is that they carry
orbital angular momentum (OAM). This was first realized by Allen and
coworkers~\cite{Allen:1992zr} for the important instance of
Laguerre-Gauss (LG) laser modes. Furthermore, they demonstrated that
these modes carry an OAM of $\ell\hbar$ per photon along the
propagation direction.

A useful feature of optical OAM is that it can be easily
manipulated and transferred; this has opened new horizons in various
fields, ranging from mechanical
micro-manipulation~\cite{Padgett:2010tg} to imaging
sciences~\cite{Maurer:2011oj,Uribe-Patarroyo:2013dw}, as well as
potential astronomical~\cite{Elias:2008hc,Tamburini:2011qt} and
communication applications~\cite{Wang:2012bs}. Beyond optical
wavelengths, OAM now plays a major role in
electron~\cite{Uchida:2010uq,Verbeeck:2010qd,
  McMorran:2011cl,karimi2012spin}, x-ray~\cite{Peele:2002wd,
  Nugent:2009vs,Hemsing:2013li} and radio frequency
engineering~\cite{Thide:2007hb,Mohammadi:2010km,Tamburini:2012vp}.

The core observation that individual photons also carry OAM brings the
most exciting possibilities for employing this variable in the quantum
regime, and a number of uses has already been
demonstrated~\cite{Mair:2001fv,Molina:2004dz,Oemrawsingh:2004fu,
Marrucci:2011we,Molina:2007qa,Fickler:2012bs}

Despite this intense activity, very little attention has been paid
thus far to the radial index $p$ of the LG modes. Usually, it is
stated that for $p > 0$, the modes are multiringed with $p + 1$
radial nodes. Beyond this short mention, no physical meaning is
attached to this quantity. Two recent papers, however, have presented
challenging and interesting insights into this
issue~\cite{Karimi:2012yt,Plick2013:sf}. Our purpose here is to
present a simple comprehensive analysis of this variable.

The two aforementioned papers considered optical modes, governed by
the paraxial wave equation. These modes are ultimately suitably
rescaled wave functions of the stationary states of a two-dimensional
quantum oscillator, under the Schr\"{o}dinger
equation~\cite{Nienhuis:2004}. Since this latter system can properly
model other interesting vortices arising in different media (as in
plasmas~\cite{Mikhailovskii:1987ys},
superfluids~\cite{Salomaa:1987ly}, and Bose-Einstein
condensates~\cite{Salomaa:1987ly}), the oscillator will serve as our
thread, bearing in mind that the results can be immediately translated
to the optical case.

\section{Stationary states of a two-dimensional oscillator} 

\subsection{Cartesian coordinates}

To be as self-contained as possible, we briefly review the example of
an isotropic two-dimensional quantum harmonic oscillator of mass $m$
and natural frequency $\omega$, with coordinates in two orthogonal
axes, say $x$ and $y$~\cite{Dodonov:1989lq,Cohen:2006bh}.  The Hamiltonian can be
compactly written as $\op{H} = \hbar \omega ( \op{n} +
\op{\openone})$, where the total number operator $\op{n}$ is
\begin{equation}
  \label{eq:Ndef}
  \op{n} = \op{n}_{x} + \op{n}_{y} = 
  \op{a}_{x}^{\dagger} \op{a}_{x} + \op{a}_{y}^{\dagger} \op{a}_{y}  \, ,
\end{equation}
and the annihilation and creation operators fulfil the canonical
commutation relations $[ \op{a}_{j}, \op{a}_{k}^{\dagger} ]
=\delta_{jk} \op{\openone}$, with $j, k\in \{x,y\}$. Since the spectrum
of $\op{n}_{j}$ is composed of all non-negative integers $n_{j}$, the
energies are given by $E_{n_{x}, n_{y}} = \hbar \omega
(n_{x} + n_{y} + 1)$ and these eigenvalues are $(n_{x} + n_{y} + 1)$-fold
degenerate.

Elements of the Fock basis are the common eigenvectors of $\op{n}_{x}$
and $\op{n}_{y}$:
\begin{equation}
  \label{eq:4}
  | n_{x},  n_{y} \rangle = \frac{1}{\sqrt{n_{x} ! n_{y} !}}  
  ( \op{a}_{x}^{\dagger} )^{n_{x}} (\op{a}_{y}^{\dagger}) ^{n_{y}} 
  |0, 0 \rangle    \, , 
\end{equation}
where $|0, 0 \rangle$ is the ground state.  The stationary states of
the oscillator are the product of Hermite-Gauss modes, as the
oscillations in each axes are kinematically independent:
\begin{eqnarray}
  \label{eq:5}
  \Psi_{n_{x}  n_{y}} (x, y) & =  & \sqrt{\frac{\alpha^{2}}{\pi \; 2^{n_{x} + n_{y}} n_{x}! n_{y}!}}
  H_{n_{x}} (\alpha x) H_{n_{y}} ( \alpha y) \nonumber \\ 
  & \times & \exp [-  \alpha^{2} (x^2+y^2)/2 ] \, , 
\end{eqnarray}
with $\alpha = \sqrt{m\omega/\hbar}$. To recover the equivalent beam
solutions, one needs to take $\alpha = \sqrt{2}$, since the paraxial
wave equation (in adimensional coordinates) coincides with the
Schr\"{o}dinger equation for the oscillator when $m=2\hbar$ and
$\omega=1$.

For our purposes, the solution at $t=0$ is enough. The wave function at
any other time can be obtained in a simple way by using the explicit
form of the propagator.  For beams, where the role of time is played
by the coordinate $z$ along the symmetry axis, this propagation brings
about additional interesting points, such as the Gouy phase.

\subsection{Cylindrical coordinates}

The axes $x$ and $y$ do not enjoy a privileged role in the
problem. Since the energy is invariant under rotations in the $xy$
plane, we could as well have chosen any other rotated reference
frame. To take a better advantage of this symmetry, we consider the
$z$-component of the angular momentum, $ \op{L}_{z} = \hbar
\op{\ell}$, with $\op{\ell} = i (\op{a}_{y}^\dagger\op{a}_{x}
-\op{a}_{x}^\dagger \op{a}_{y} )$, and use the rotated bosonic
operators~\cite{Sanchez-Soto:2013gf}
\begin{equation}
  \label{eq:7}
  \op{a}_{\pm}= \frac{1}{\sqrt{2}}  ( \op{a}_{x} \mp i  \op{a}_{y}) \, , 
  \qquad 
  \op{a}_{\pm}^{\dagger} = \frac{1}{\sqrt{2}} 
  ( \op{a}_{x}^\dagger \pm i  \op{a}_{y}^\dagger) \, ,
\end{equation}
where $[\op{a}_{j},\op{a}_{k}^{\dagger}] =\delta_{jk} \op{\openone}$,
with $j, k \in \{+,-\}$. We can then check that
\begin{equation}
  \label{eq:8}
  \op{n} =  \op{n}_{+} + \op{n}_{-}  \, , 
  \qquad
  \op{\ell} = \op{n}_{+} - \op{n}_{-} \, ,
\end{equation} 
whose interpretation is direct: the system can be envisioned now as
consisting of ``quanta'' with positive (counterclockwise rotation
around $z$) and negative (clockwise rotation around $z$) orbital
angular momentum.

The Fock basis $\{ |n_{+}, n_{-} \rangle \}$ of the common
eigenvectors of $\op{n}_{+}$ and $\op{n}_{-}$ can be constructed much
in the same way as in Eq.~(\ref{eq:4}).  However, it will prove useful
to consider instead the continuous set
\begin{equation}
  \label{eq:11}
  | \eta  \rangle =  
  \exp \left (  
    -\frac{1}{2} |\eta |^2 +
    \eta  \op{a}_{+}^{\dagger}  -\eta ^{\ast} \op{a}_{-}^{\dagger} +
    \op{a}_{+}^{\dagger} \op{a}_{-}^{\dagger} 
  \right ) |0,0\rangle \, ,
\end{equation}
parametrized by the complex number $\eta = r \exp (- i \varphi )$. The
states $|\eta \rangle$ constitute an orthonormal basis, whose
properties have been reviewed in depth in Ref.~\cite{Hong-yi:1994rw}.
In the representation they generate [$\psi (\eta ) = \langle \eta |
\psi \rangle$], the action of the basic operators is
\begin{eqnarray}
  \label{eq:12}
  \op{a}_{+} \psi (\eta  ) & = & \left ( \frac{\eta}{2} +
    \frac{\partial}{\partial \eta^{\ast}} \right ) \psi (\eta ) \, ,
  \nonumber \\
& & \\
  \op{a}_{-} \psi (\eta  ) & = & -  \left ( \frac{\eta^{\ast}}{2} +
    \frac{\partial}{\partial \eta} \right ) \psi(\eta ) \, , \nonumber
\end{eqnarray}
while for the adjoints we have
\begin{equation}
  \label{eq:adj}
  \op{a}_{+}^{\dagger} = \op{a}_{-} - \eta^{\ast} \, , 
  \qquad
  \op{a}_{-}^{\dagger} = \op{a}_{+} - \eta \, . 
\end{equation}
Since the exponential acting on the vacuum in Eq.~\eqref{eq:11} is not
unitary, the creation and destruction operators are not conjugates one
of the other under the usual boson conjugation.

Given the above, $\op{n}$ and $\op{\ell}$ act in this space as
\begin{eqnarray}
  \label{eq:NLdif}
  \op{n} & \mapsto &  \frac{r^2}{2} - 
    \frac{1}{2} \left (\frac{\partial^{2}}{\partial r^{2}} +
      \frac{1}{r}  \frac{\partial}{\partial r} +
      \frac{1}{r^2}\frac{\partial^{2}}{\partial \varphi^{2}} \right )
    - 1 \, , \nonumber \\
  & & \\
  \op{\ell} & \mapsto & - i \frac{\partial}{\partial \varphi} \, . 
\nonumber
\end{eqnarray}

As $[\op{\ell}, \op{n} ] = 0$, the basis $\{ |n_{+},n_{-} \rangle \}$ can
be reinterpreted as common eigenvectors of $\op{n}$ and
$\op{\ell}$, with eigenvalues $n= n_{+} +  n_{-}$ and $\ell = n_{+} -
n_{-}$, respectively. The stationary states in this basis can be
readily obtained using Eqs.~(\ref{eq:NLdif}); the final result is
\begin{equation}
  \label{eq:14}
  \Psi_{n \ell} (r, \varphi ) = 
  A_{n \ell} (r) \, e^{i\ell \varphi} \, ,
\end{equation}
where the normalized amplitude is 
\begin{equation}
  \label{eq:amp}
  A_{n \ell} (r) = \frac{\sqrt{2\alpha^{2} p!}}{\sqrt{(p + |\ell |) !}} \,  
  e^{- \alpha^{2} r^2/2} (\alpha r)^{|\ell|}
  L^{|\ell |}_{p} (\alpha^{2} r^2) \, ,
\end{equation}
$L_{p}^{\ell} (x)$ are the generalized Laguerre polynomials and
we have written $p = (n - |\ell |)/2$. The probability distribution $|
\Psi_{n \ell} (r, \varphi)|^{2}$ shows $p$ dark concentric rings.

\section{Quantum optics with radial modes}

\subsection{The radial number operator} 

Since the number of dark rings is $p= (n-|\ell|)/2$, the operator
\begin{equation}
  \label{eq:defP}
  \op{p}  = \frac{1}{2} (\op{n}- | \op{\ell} |) = \left \{ 
    \begin{array}{ll}
      \op{n}_{-}  \quad & \mathrm{for} \ \ell > 0  \, , \\
      & \\
      \op{n}_{+}  & \mathrm{for} \ \ell < 0 \, ,
    \end{array}
  \right .
\end{equation}
seems to be a sensible definition for the radial-number operator of
the Laguerre-Gauss modes.  According to Eq.~(\ref{eq:NLdif}), in
differential form it reads
\begin{equation}
  \label{eq:15}
  \op{p}  \mapsto  - \frac{1}{4}  
  \left(\pdd{}{r}+\frac{1}{r}\pd{}{r}+\frac{1}{r^2}
    \pdd{}{\varphi}\right)+\frac{i}{2}\pd{}{\varphi}+\frac{1}{2}\frac{r^2}{2}-
  \frac{1}{2} \,. 
\end{equation}
Incidentally, it coincides with the operator found in
Ref.~\cite{Plick2013:sf} by setting $r \rightarrow\alpha r$.

To simplify the following discussion, let us, for the time being,
relabel the stationary states $|n, \ell \rangle$ as $|p, \ell 
\rangle$, where $p$ indicates the radial mode eigenvalue; i.e., 
\begin{equation}
  \label{eq:16}
  \op{p} | p, \ell \rangle = p | p, \ell \rangle \, . 
\end{equation}
As heralded in the Introduction, we are interested in exploring the
Hilbert space associated with the radial number $p$, while keeping the
OAM $\ell$ fixed. At first sight, one might
look for the canonical conjugate variable to $\op{p}$. Since,
according to Eq.~(\ref{eq:defP}), $\op{p} = \op{n}_{-}$ or
$\op{n}_{+}$ (depending on the sign of $\ell$), such a variable would
be a phase $\op{\phi}_{-}$ or $\op{\phi}_{+}$. This means that if we
denote by $\op{e}= \exp (i\op{\phi}_{\pm})$ the exponential of such a
putative phase, the corresponding commutation relation will
read~\cite{Lynch:1995cq}
\begin{equation}
  \label{eq:PE}
  [ \op{e}, \op{p} ] = \op{e} \, .
\end{equation}
This immediately implies
\begin{equation}
  \label{eq:elad}
  \op{e} = \sum_{p=0}^{\infty} |p, \ell \rangle \langle p+1, \ell | \, ,
 \end{equation}
so that
\begin{equation}
  \label{eq:elad}
  \op{e} |p, \ell \rangle = |  p-1 , \ell \rangle \, ,
  \qquad
  \op{e}^{\dagger}  |p, \ell \rangle = |  p + 1 , \ell \rangle \, .
\end{equation}
Whereas the spectrum of $\op{\ell}$ is unbounded, including all the
integer numbers, the spectrum of $\op{p}$ is semibounded, as it
comprises only  non-negative integers. This indicates that the action of
$\op{e}$ as a ladder operator fails at $p=0$, and consequently it
cannot be unitary:
\begin{equation}
  \label{eq:enonu}
  \op{e} \op{e} ^{\dagger} = \op{\openone} \, , 
  \qquad
  \op{e} ^{\dagger} \op{e} =  \op{\openone} - \op{\mathcal{P}}_{0} \, ,
\end{equation}
where $\op{\mathcal{P}}_{0} = |0, \ell \rangle \langle 0, \ell |$ is
the projector on the ``vacuum.''

All these problems thus place this interpretation on shaky grounds.
For this reason, we prefer to follow an alternative route. To this
end, we observe that to increase ( decrease) the radial number by
one unit, with $\ell$ unchanged, we need to create (annihilate)
one positive quantum and one negative quantum, namely
\begin{eqnarray}
  \label{eq:17}
  \op{k}_{+} |p, \ell \rangle & =  & 
  \op{a}_{-}^\dagger \op{a}_{+}^\dagger |p,\ell\rangle 
  \propto |p+1,\ell \rangle  \, , \nonumber \\
  & & \\
  \op{k}_{-} |p,\ell\rangle & = & \op{a}_{-} \op{a}_{+} |p,\ell\rangle
  \propto |p-1,\ell\rangle \, . \nonumber 
\end{eqnarray}
One can check that
\begin{equation}
  \label{eq:ccr}
  [\op{k}_{+} , \op{k}_{-} ]  =  - 2 \op{k}_{z} \, ,
  \quad
  [\op{k}_{z} ,\op{k}_{+} ]=\op{k}_{+} \, ,
  \quad
  [\op{k}_{z} ,\op{k}_{-} ] =- \op{k}_{-} \, , 
\end{equation}
with $\op{k}_{z}= (\op{n} + \op{\openone} )/2$. This means that if we
define $\op{k}_{\pm} = \op{k}_{x} \pm i \op{k}_{y}$, we have 
\begin{equation}
  \label{eq:18}
  [\op{k}_{x} , \op{k}_{y} ]  =  - i \op{k}_{z} \, ,
  \quad
  [\op{k}_{y} ,\op{k}_{z} ] = i \op{k}_{x} \, ,
  \quad
  [\op{k}_{z} ,\op{k}_{x} ]= i \op{k}_{y} \, ,
\end{equation}
that is, they are the generators of the su(1, 1) algebra, as first
noticed in Ref.~\cite{Karimi:2012yt}.

\subsection{Radial coherent states} 

To explore the issue in more detail, it is convenient to give some
basic background on some well-known irreducible representations (irreps)
of SU(1,1), which are excellently reviewed in
Ref.~\cite{Biedenharn:1965it} and whose role in quantum optics is
difficult to underestimate~\cite{Gerry:1985fk,Yurke:1986yg,
Buzek:1989wd,Vourdas:1992vl,Ban:1993pi,Brif:1996oj,Gerry:1997kn,
Agarwal:2001dw,Mias:2008cr,Marino:2012ct}.

The Casimir operator for this group is $\op{K}^{2} = \op{k}_{z}^{2} -
\op{k}_{x}^{2} - \op{k}_{y}^{2}$, which can be expressed as
$\op{K}^{2} = k (k-1) \op{\openone}$, where the Bargmann index $k$
labels the different irreps (this index plays the role of
spin for rotations). In our case, a simple calculation shows that $k =
(| \ell | + 1)/2$, so that $k = 1/2, 1, 3/2, \ldots$, which
corresponds to the so-called positive discrete series, for which $
\op{k}_{z}$ is diagonal and has a discrete spectrum. In the Fock basis
$\{ |n_{+}, n_{-} \rangle \}$, the basis states of the irrep $k$ are 
$ \{ | k , k + p \rangle \}$, with $p=0, 1, \ldots$, and hence
\begin{eqnarray}
  \label{eq:diag}
  \op{k}_{z} |k, k+p \rangle  =   
  (k+p)  |k, k+ p \rangle \, , 
\end{eqnarray}
while the ladder operators act as
\begin{eqnarray}
  \label{eq:ladd}
  \op{k}_{+} |k, k+p \rangle  & =  & 
  \sqrt{(2k+p) (p+1)}  |k, k+ p +1 \rangle \, , 
  \nonumber \\
  & & \\
  \op{k}_{-} |k, k+p \rangle & =  &
  \sqrt{p (2k+p- 1 )}  |k, k+p - 1 \rangle \, . 
  \nonumber 
\end{eqnarray}
Note that we can make the identification $ |p,\ell\rangle
\leftrightarrow |k, k+ p \rangle$, provided $k = (| \ell | + 1)/2$.

Since $\op{k}_{-} |k,k \rangle = 0$, this state can be taken as the
vacuum. Indeed, $\op{D} (\xi) = \exp ( \xi \op{k}_{+} - \xi^{\ast}
\op{k}_{-} )$ are truly displacement operators, so according to
Perelomov prescription~\cite{Perelomov:1986kl}, the set
\begin{equation}
  \label{eq:21}
  | \zeta \rangle =  \op{D} (\xi) |k, k \rangle
\end{equation}
constitutes a family of \emph{bona fide} coherent states parametrized
by the pseudo-Euclidean unit vector $\mathbf{n} = (\sinh \omega \cos
\varphi, \sinh \omega \sin \varphi, \cosh \omega )$, with $\xi =
(\omega /2) \exp (i \varphi) $ and $\zeta = \tanh (\omega /2) \exp (-i
\varphi)$.

By expanding the exponential and employing the disentangling theorem, we
get the decomposition
\begin{equation}
  \label{eq:6}
  | \zeta \rangle = ( 1 - |\zeta |^{2})^{k} \sum_{p=0}^{\infty}
  \sqrt{\frac{\Gamma (2k + p)}{p! \Gamma (2k)}} \zeta^{p} 
  |k, k + p \rangle \, , 
\end{equation}
and by projecting over the complete basis $| \eta \rangle$ we get the
corresponding wave function $\Psi_{\zeta} (r, \varphi)$ in the
transverse parameters. The expression can be simplified into an
exponential form using the identity
\begin{equation}
  \exp { \left(\frac{\gamma x}{\gamma -1}\right) } = 
  (1-\gamma)^{1+|\ell|} \sum_{p=0}^{\infty} {\gamma}^{p}
  L_{p}^{|\ell|}(x) \, ,
\end{equation} 
so the final result is
\begin{equation}
  \label{eq:rcs}
  \Psi_{\zeta} ( r , \varphi) =\sqrt{\frac{\alpha^{2}}{\pi|\ell|!}}
  \left [ \frac{1-|\zeta|^2}{(1-\zeta)^2}\right]^{\frac{|\ell|+1}{2}} \!\!
  e^{\frac{\zeta+1}{\zeta -1} \,\alpha^{2} r^2/2}  (\alpha r)^{|\ell|}
  e^{i \ell\varphi} \, .
\end{equation} 
We see that the Perelomov coherent states are polynomial-Gauss modes
at $t=0$; a subfamily of Hypergeometric-Gauss modes upon
evolution, as discussed in Ref.~\cite{Karimi:2007fj}. They are also
eigenstates of the OAM, and shape invariant in the time evolution.

\begin{figure}
 \includegraphics[width=0.80\columnwidth]{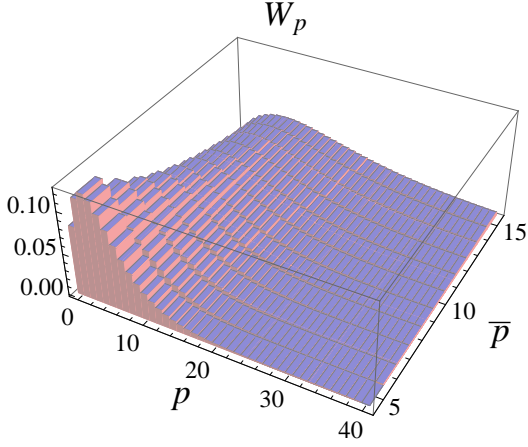}
  \caption{(Color online). Probability distribution $W_{p}$ for a
    coherent state  $| \zeta \rangle $ with $\ell =1$ as as a function
  of $p$ and the ring average number  $\bar{p}$.}
  \label{fig:dist}
\end{figure}

\begin{figure}
  \includegraphics[width=0.92\columnwidth]{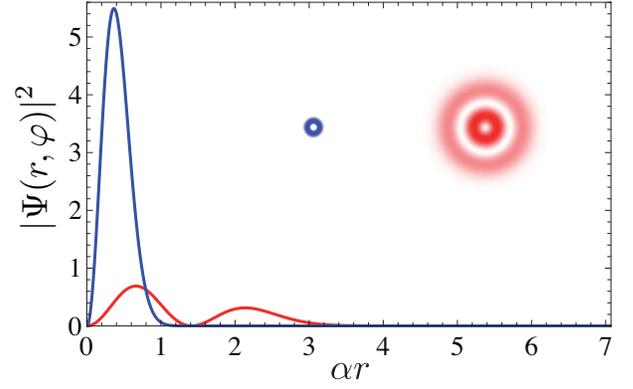}
  \caption{(Color online). Intensity profiles $| \Psi (r, \phi)|^{2}$
    for a radial coherent state $ | \zeta \rangle$ written as in
    \eqref{eq:rcs}, with $\langle p \rangle = 1$ and for an eigenstate
    $|p , \ell \rangle$ with $p=1$ and $\ell =1$. In the inset, we
    show the corresponding density plots, in the same order.}
  \label{fig:Trans1}
\end{figure}

The average number of sharp rings in the state $ | \zeta \rangle$ is 
\begin{equation}
  \label{eq:9}
  \bar{p} = \frac{| \zeta |^{2}}{|\zeta|^{2} - 1} (| \ell | + 1) \, ,
\end{equation}
and the statistical distribution of rings $W_{p} =| \langle p,\ell |
\zeta \rangle |^{2} $ is
\begin{eqnarray}
  \label{eq:13}
  W_{p}  =  \frac{(| \ell | +1)^{|\ell| + 1} (p + |\ell|)!}{p! | \ell |!}  
\frac{ \bar{p}^{p}}{(\bar{p} +|\ell| +1)^{p + |\ell| +1}} \, . 
\end{eqnarray}
In Fig.~1 we have plotted this distribution for a coherent state with
$\ell =1$ and different values of $\bar{p}$. In Fig.~\ref{fig:Trans1},
we have plotted $| \Psi_{\zeta} (r, \varphi)|^{2}$ for a coherent
state with $\ell =1$ and $\langle \op{p} \rangle = 1$, and the
corresponding distribution for the Laguerre-Gauss eigenstate with
$p=1$. The striking differences can be appreciated at a glance.

It is worth mentioning that there is an alternative
definition of coherent states, due to Barut and
Girardello~\cite{Barut:1971yq}:
\begin{equation}
  \label{eq:BGdef}
  \op{k}_{-} | \zeta \rangle_{\mathrm{BG}} = \zeta | \zeta
  \rangle_{\mathrm{BG}}\, ,
\end{equation}
which appears as a reasonable generalization of the standard coherent
states as eigenstates of the annihilation operator. This equation can
be solved in the $|p, \ell \rangle $ basis, yielding
\begin{equation}
  \label{eq:BGst}
  | \zeta \rangle_{\mathrm{BG}} = \frac{| \zeta |^{\ell /2}}
 {\sqrt{I_{\ell} (2 |\zeta|)}} 
\sum_{p=0}^{\infty}  
\frac{\zeta^{p}} {\sqrt{p! (p + |\ell |)!}} 
|p, \ell \rangle \, ,
\end{equation}
where $I_{\ell} (x)$ is the modified Bessel function. Projecting again in
the transverse coordinates we get, after some calculations,
\begin{equation}
  \label{eq:BGproj}
 \Psi_{\zeta,{\mathrm{BG}}} (r, \varphi) = \sqrt{\frac{\alpha^{2}}{\pi I_{| \ell |} ( 2 |\zeta |^{2})}}
e^{(\zeta^{2} - \alpha^{2} r^{2}/2)} 
 J_{| \ell |} (  2 \zeta r) e^{i \ell   \varphi} \, .
\end{equation}
This set of coherent states are thus realized as Bessel-Gauss
modes. However, these solutions are not shape invariant, which runs
against the notion of coherence.

\subsection{Radial intelligent and squeezed states}

The commutation relations (\ref{eq:ccr}) imply that these
operators cannot be measured simultaneously, which is reflected by the
uncertainty relation
\begin{equation}
  \label{eq:ur}
  \Delta \op{k}_{x} \, \Delta \op{k}_{y} \ge \textstyle{\frac{1}{2}} 
| \langle \op{k}_{z} \rangle | \, ,
\end{equation}
where $\Delta \op{A} = [ \langle \op{A}^{2}
\rangle - \langle \op{A} \rangle^{2} ]^{1/2}$ stands for the
variance. According to the standard definition, squeezing
occurs whenever~\cite{Wodkiewicz:1985qf}
\begin{equation}
  \label{eq:10}
  (\Delta \op{k}_{x})^{2} \le \textstyle{\frac{1}{2}} 
  |\langle  \op{k}_{z} \rangle |
  \qquad
  \mathrm{or}
  \quad
  (\Delta \op{k}_{y})^{2} \le \textstyle{\frac{1}{2}} 
  |\langle  \op{k}_{z} \rangle |\, .
\end{equation}
Intelligent states are those for which (\ref{eq:ur}) holds as an
equality. The coherent states (\ref{eq:rcs}) and (\ref{eq:BGproj})
are intelligent but not squeezed.  

Indeed, these intelligent states are solutions of the eigenvalue
problem~\cite{Jackiw:1968gd}  
\begin{equation}
  \label{eq:defIS}
  (\op{k}_{x} - i \lambda \op{k}_{y} ) | \Psi_{\lambda } \rangle = 
\Lambda |\Psi_{\lambda}  \rangle \, ,  
\qquad
\lambda \in \mathbb{R} \, .
\end{equation}
Although they have been investigated from various
perspectives~\cite{Bergou:1991km,Gerry:1995kq,Puri:1996qm}, we follow
here the comprehensive approach of Ref.~\cite{Joanis:2010yq}, which
starts by noting that the coherent state $\exp ( i \tau \op{k}_{y} )
|k , k \rangle$ is intelligent provided $\lambda= \cosh \tau$ [with
eigenvalue $\Lambda = - (k + M) \sinh \tau$, and $M = 0, 1, \ldots$ an
integer number]. Then, the most general intelligent state can be
written as
\begin{equation}
  \label{eq:puf}
  | \Psi_{\ell, M} (\tau) \rangle = \exp (  i \tau \op{k}_{y} ) 
  | \kappa_{M}^{k}  (\tau ) \rangle \, ,
\end{equation} 
where $\tau$ is the squeezing parameter and the seed state
$|\kappa_{M}^{k} \rangle$ can be expressed as 
\begin{equation}
  \label{eq:2}
  | \kappa_{M}^{k} (\tau ) \rangle = \sum_{p=0}^{M} 
c_{p}^{k} (\tau ) |k, k + p \rangle \, .
\end{equation}
The coefficients $c_{p}^{k}$ can be obtained as a recursion relation;
the final result reads 
\begin{equation}
  \label{eq:1}
c_{p}^{k} =\binom{M}{p} 
\displaystyle
\frac{\tanh^{p} \tau}{\binom{2 k+ p-1}{p}^{{1/2}}}
\, c_{0}^{k}\, ,
\end{equation}
and $c_{0}^{k}$ is fixed by the normalization of the state. The infinite family
(\ref{eq:puf}) of states parametrized by $k$ and $M$ is actually squeezed.

\begin{figure}
  \includegraphics[width=0.92\columnwidth]{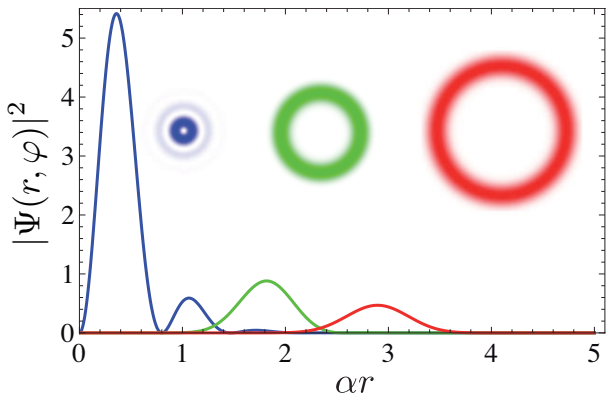}
  \caption{(Color online). Intensity profiles $| \Psi
    (r,\varphi)|^{2}$ for intelligent states with $M=10$ and $\tau =
    1/2$ for several values of $\ell$: $\ell = 1$, $\ell=10$, and
    $\ell=20$. The inset shows the corresponding density plots, in the
    same order.}
  \label{fig:variableell}
\end{figure}

\begin{figure}[t]
  \includegraphics[scale=0.6]{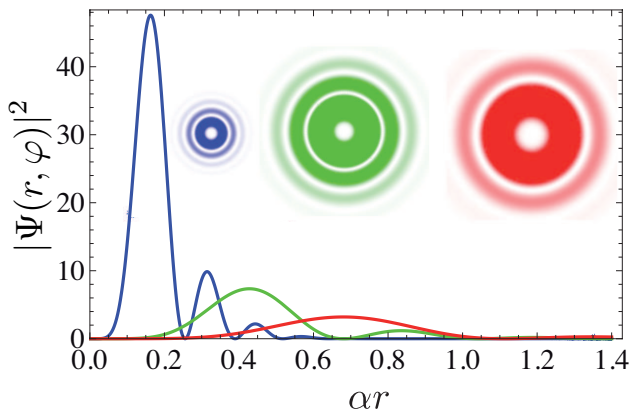}
  \caption{(Color online). Intensity profiles $| \Psi
    (r,\varphi)|^{2}$ for intelligent states with $M=11$ and $\ell =3$
    for several values of $\tau$: $\tau = 16/5$, $\tau=13/10$, and
    $\tau =3/5$. The inset shows the corresponding density plots,
    in the same order.}
  \label{fig:FFplot3D24}
\end{figure}

Next, we need to project $\exp ( i \tau \op{k}_{y} )|\kappa_{M}^{k}
(\tau ) \rangle$ on the basis $| k , k + p \rangle $. To this end, we
recall that the action of $\exp ( i \tau \op{k}_{y} ) $ on a basis
state $|k, k + p\rangle$ is given in terms of SU(1,1) Wigner 
$d$-functions~\cite{Ui:1970nh} 
\begin{equation}
\exp ( i \tau \op{k}_{y} )|k, k + p\rangle
=\sum_{p^{\prime}=0}^\infty  d^k_{k+p,k+p^{\prime}} (- \tau ) \, 
|k, k +p^{\prime}\rangle \, .
\end{equation} 
In this way, we get,  expressed  in transverse coordinates
\begin{equation}
\Psi_{\ell M} (r, \varphi, \tau )= \sum_{p^{\prime}=0}^{\infty} 
 \sum_{p=0}^{M}  d^k_{k+p^{\prime},k+p} (- \tau) \, c_{p}^{k} (\tau ) \, 
 A_{p \ 2k-1 }(r) e^{i\ell\varphi} \, ,
\end{equation} 
where $A_{p \ell}$ has been defined in Eq.~\eqref{eq:amp}.
The effect of increasing $\ell$, for fixed $p$ and $\tau$, is
illustrated by plotting the intensity profile $|
\Psi_{\ell,M}(r,\varphi, \tau) |^2$ as a function of $ r$ in
Fig.~\ref{fig:variableell}: this intensity tends
to a Gaussian-like shape. The effect of increasing $\tau$ for fixed
$\ell$ and $M$ is  illustrated in Fig.~\ref{fig:FFplot3D24}, and leads
to the appearance of rings as we increase $\tau$.

For large values of $\ell$ (more concretely, for  $k$ and  $p$ large,
but $p/k\ll 1$), one has at hand a compact asymptotic approximation
to the $d$ functions, namely~\cite{Rowe:2001uq} 
\begin{equation}
  d^{k}_{k+p,k}(\tau) \simeq \frac{1}
  {[ (k+p)^2-k^2 ]^{1/4}} \,e^{-k(\tau-\tau_p)^2/2}\, ,
\end{equation}
with $\cosh \tau_{p}=(k+p)/k$ and whose Gaussian nature is evident.

\section{Concluding remarks} 

In summary, we have provided a  handy toolbox to deal with the radial
index of Laguerre-Gauss modes and shown how it can be
used to construct a consistent quantum theory of this variable. We
stress that this is more than an academic curiosity, since recent
experiments in our laboratory~\cite{Karimi:2014by} have confirmed that
the radial degree of freedom of single  photons can be manipulated
individually in a quantum regime.  It is our hope that the results
presented here will inspire novel quantum protocols and algorithms
using  such a ``forgotten quantum number''.

\begin{acknowledgments}
  E. K. and R. W. B. acknowledge the support of the Canada Excellence
  Research Chairs (CERC) Program. The work of H. G. is supported by
  the Natural Sciences and Engineering Research Council (NSERC) of
  Canada. J. R. and Z. H. are grateful to the financial assistance of the
  Technology Agency of the Czech Republic (Grant TE01020229) and the
  Czech Ministry of Industry and Trade (Grant FR-TI1/364). G. L. is
  partially funded by EU FP7 (Grant Q-ESSENCE).  Finally, P. H. and
  L. L. S. S. acknowledge the support from the Spanish MINECO (Grant
  FIS2011-26786).
\end{acknowledgments}


%

\end{document}